\documentclass[12pt]{article}
\usepackage{natbib,url,graphicx,setspace}
\newtheorem{theorem}{Theorem}
\newtheorem{lemma}{Lemma}
\begin{document}

\title{Empirical Likelihood Confidence Intervals for Nonparametric Functional Data Analysis }         
\author{Heng Lian\\Division of Mathematical Sciences\\School of Physical and Mathematical Sciences\\Nanyang Technological University\\Singapore, 637371\\Singapore}        
\date{}          
\maketitle
\section*{Abstract}
We consider the problem of constructing confidence intervals for nonparametric functional data analysis using empirical likelihood. In this doubly infinite-dimensional context, we demonstrate the Wilks's phenomenon and propose a bias-corrected construction that requires neither undersmoothing nor direct bias estimation. We also extend our results to partially linear regression involving functional data. Our numerical results demonstrated the improved performance of empirical likelihood over approximation based on asymptotic normality. 

\noindent\textbf{Keywords:} Empirical likelihood; Confidence interval; Functional data analysis; Nonparametric regression; Wilks's theorem.
\section{Introduction}       
Recently there has been an explosion of interests in the study of functional data, where the independent variables in the statistical problem are curves, or more abstractly, elements belonging to a metric space \citep{ramsay82, ramsay91, ferraty04}. As a natural extension of multivariate data analysis, functional data analysis provides valuable insights into these problems. Compared with the discrete multivariate analysis, functional analysis takes into account the smoothness of the high-dimensional covariates, and often suggests new approaches to the problems that have not been discovered before. Even for non-functional data, the functional approach can often offer new perspectives on the old problem.

This statistical area was first developed from extending the parametric linear models to the functional context, followed by the more recent literature on nonparametric techniques, mostly focusing on the kernel regression method. For an introductory exposition on this field, we will refer the reader to the two monographs, \cite{ramsey05} and \cite{ferraty06}, for the parametric and nonparametric approach respectively. More recently, there has also been some studies using alternative nonparametric techniques such as reproducing kernel Hilbert spaces \citep{preda07,hernandez07}, and dealing with the case where the dependent variables are curves as well  \citep{cuevas02,lian07}.

We are concerned here with the nonparametric regression problem using the Nadaraya-Watson estimator. As pointed out in \cite{ferraty02a}, although the extension of the Nadaraya-Watson estimator is relatively straightforward in computational aspects, mathematical challenges arise from the fact that we are now dealing with a doubly infinite-dimensional situation: both the regression function and the covariate belong to an infinite-dimensional space. By now there have appeared many studies extending various kernel regression results to the functional case, for i.i.d as well as weakly dependent data. \cite{ferraty04} established the convergence rates for the kernel estimator, while asymptotic normality was extended to the functional context in \cite{ferraty07} and \cite{masry05} for i.i.d. and strongly mixing sequence respectively. Nearest neighbor regression appeared in \cite{burba09}. Similar techniques have also been extended to semiparametric problems such as partially linear models where the nonparametric part is estimated by the Nadaraya-Watson estimator \citep{perez06}. These studies confirmed the applicability of traditional nonparametric method in functional contexts. All of these results also demonstrated the importance of appropriately dealing with the data sparsity problem caused by the infinite dimensionality of the dependent variables, and various semi-metrics have been proposed to alleviate this problem. 

As is prevalent in the statistical community, one can argue that the assessment of uncertainty of an obtained estimator is an important step in all statistical analysis. This aspect is so far ignored in the literature for nonparametric functional data analysis. The asymptotic normality of the Nadaraya-Watson estimator shown in \cite{ferraty07} with explicit expressions for the bias and variance terms provides us with a mechanism for constructing asymptotically valid point-wise confidence intervals. But besides the fact that the expressions for bias and variance involve unknown parameters and thus need to be estimated from the data which is itself a difficult problem, our simulations also show the intervals constructed in this way have poor coverage rates for finite sample sizes. 

In this article, we propose to adapt the empirical likelihood method, first introduced by \cite{owen88}, to construct point-wise confidence intervals for the regression function. A major advantage of empirical likelihood is that it involves no predetermined assumptions on the shape of the confidence interval, while the interval constructed by normal approximation is always symmetric around the point estimator. It is proved in \cite{diciccio91} that empirical likelihood is Bartlett-correctable and thus it has an advantage over another popular nonparametric method for establishing confidence intervals, the bootstrap. The general property of empirical likelihood was studied by \cite{owen90}, and \cite{qin94} gave a much more general theory on empirical likelihood properties with estimating equations. In the context of kernel density estimation, \cite{hall93} studied the empirical likelihood confidence bands, while \cite{chen96} provided finer analysis for point-wise confidence intervals and his simulation clearly demonstrated the improved coverage accuracy of empirical likelihood over percentile-t bootstrap.

The purpose of our study is to establish the Wilks's phenomenon for empirical likelihood in nonparametric functional regression with strongly mixing data. Although empirical likelihood has been used in many different problems, the application of this method in functional data analysis is new and development its asymptotic properties is more involved due to the double infinite dimensionality problem mentioned above. It is well-known that in kernel regression, the constructed interval has a non-ignorable bias when the optimal bandwidth for function estimation is used. Similar to the intervals constructed by bootstrap, there are in general two approaches to address this problem. One is to correct the bias by undersmoothing, and the other is to use the explicit bias formula to shift the constructed intervals, with unknown quantities estimated by the data and plugged into the expression. We find that in functional data regression, the first approach using smaller bandwidth aggravates the data sparsity problem and in our simulations using a small bandwidth causes the estimation problem to be very unstable since there are very few covariates found within the smaller neighborhood of the testing covariates. For the second approach, as mentioned above, the bias is difficult to estimate as can be seen from the expression for bias which is discussed in more detail later. Thus we propose an implicit method that first uses the optimal bandwidth to obtain an estimate of the regression function and then uses the estimate to correct for bias in the estimating equation. Our simulations show that the bias-corrected intervals have  improved coverage rates.

The organization of the paper is as follows. Section 2 presents the model for empirical likelihood interval construction. We establish the Wilks's theorem for empirical likelihood with dependent data. This theoretical generality is required when we work with time series data later. Then the bias corrected interval is constructed and we also briefly discuss the extension to partially linear models which greatly expands the scope of the method. In Section 3, we use both simulated and real data to show the improved accuracy of confidence intervals constructed by empirical likelihood over those by normal approximation. Section 4 is devoted to a discussion of the results. The technical proofs are collected in the Appendix.

\section{Empirical likelihood inference for nonparametric functional analysis}
\subsection{Nonparametric functional model}
Let $\{Y_i,X_i\}_{i=1}^n$ be a stationary and ergodic process marginally distributed as $(Y,X)$. The functional nonparametric regression model, introduced in \cite{ferraty04}, is defined as
\[Y_i = r(X_i) + \epsilon_i. \] 
We assume that $\epsilon_i$ is a random variable with $E(\epsilon_i|X_i) = 0$ and
$Var(\epsilon_i|X_i) = \sigma^2(X_i)$. The dependent variable $Y_i$ is real-valued, and the covariate $X_i$ is assumed to belong to some semi-metric vectorial space $H$ equipped with a semi-metric $d(., .)$.

Estimation of the regression function $r$ is a crucial issue in nonparametric functional model and \cite{ferraty04} proposed the adaptation of Nadaraya-Watson estimator to the functional context for a fixed $x_0$:
\[\hat{r}(x_0)=\frac{\sum_{i=1}^n K(d(X_i,x_0)/h)Y_i}{\sum_{i=1}^n K(d(X_i,x_0)/h)},\]
where $K$ is the kernel and $h$ is the bandwidth, which satisfies $h\rightarrow 0$ as $n\rightarrow\infty$.

The asymptotic properties of $\hat{r}$ crucially depends on the following so-called small ball probabilities on the covariates, since these probabilities are directly related to the data sparsity problem that often plagues the infinite-dimensional model:
\[\phi(h)=P(X\in B(x_0,h)),\]
where $B(x_0,h)=\{x\in H, d(x,x_0)\le h\}$ is a neighborhood of $x_0$. Note that since we always consider a fixed $x_0$ for estimation throughout, we will omit the dependence of $\phi(h)$ on $x_0$ in our notation. For weakly dependent data, the properties of the estimator also depend on the pairwise joint distribution of the covariates, and we define
\[\psi(h)=\sup_{i\neq j}P(X_i\in B(x_0,h), X_j\in B(x_0,h)).\]
Note for independent data sequence, we obviously have $\psi(h)=\phi(h)^2$.

For weakly dependent sequence data as dealt with in our current paper, rates of almost complete convergence (slightly stronger than almost sure convergence) were obtained in \cite{ferraty04,ferraty06}. Later works by \cite{masry05,ferraty07} showed that $\sqrt{n\phi(h)}(\hat{r}(x_0)-r(x_0)-b(x_0))$ is asymptotically normal under suitable conditions. Using smaller $h$, the bias term $b(x_0)$ can disappear at the cost of enlarged asymptotic variance.

\subsection{Empirical likelihood inferences}
There is an estimating equation representation for the Nadaraya-Watson estimator. The estimator is actually the solution to the equation
\[\sum_{i=1}^nK_i(Y_i-\mu)=0,\]
where we use $K_i=K(d(X_i,x_0)/h)$ to simplify the notation. The corresponding population version is 
\[EK(Y-\mu)=0,\]
where $K=K(d(X,x_0)/h)$. Obviously the zero of the above equation is $\mu_0=EKY/EK$ instead of $r(x_0)$. Note that $\mu_0$ implicitly depends on $h$ which in turn depends on $n$. 

Now we can define a likelihood function for $\mu$ based on the empirical likelihood principle as
\[L_n(\mu)=\max\left\{\prod_{i=1}^n p_i|\,p_i\ge 0, \sum_{i=1}^np_i=1, \sum_{i=1}^np_iK_i(Y_i-\mu)=0\right\},\]
where we set $L_n(\mu)=0$ if there is no $\{p_i\}_{i=1}^n$ satisfying the above constraints. It is easily seen that the maximum empirical likelihood estimator coincides with the Nadaraya-Watson estimator $\hat{r}(x_0)$ with $L_n(\hat{r}(x_0))=n^{-n}$, achieved when $p_i=1/n$. The corresponding nonparametric likelihood ratio is given by
\[LR_n(\mu)=L_n(\mu)/L_n(\hat{r}(x_0))=\max\left\{\prod_{i=1}^n (np_i)|\,p_i\ge 0, \sum_{i=1}^np_i=1, \sum_{i=1}^np_iK_i(Y_i-\mu)=0\right\}.\]

Using the duality approach, the log likelihood ratio (multiplied by a constant $-2$) becomes
\[lr_n(\mu)=-2\log LR_n(\mu)=2\sum_{i=1}^n(1+\lambda K_i(Y_i-\mu)),\]
where the Lagrange multiplier $\lambda$ solves
\[\sum_{i=1}^n\frac{K_i(Y_i-\mu)}{1+\lambda K_i(Y_i-\mu)}=0.\]

Now we give the first result of the paper. 
\begin{theorem}\label{th:1}
Suppose that conditions (c0)-(c8) in the Appendix hold, then 

(a) $lr_n(\mu_0)\rightarrow \chi_1^2$ in distribution

(b) If in addition $nh^{2\alpha}\phi(h)\rightarrow 0$, then $lr_n(r(x_0))\rightarrow \chi_1^2$ in distribution, where $\alpha$ is defined in condition (c2) in the Appendix.
\end{theorem}

Thus, according to part (a) of the theorem, we can use the empirical likelihood ratio to construct a $1-\alpha$ confidence interval for $\mu_0$ as 
\[ \{\mu:lr_n(\mu)\le\chi^2_1(\alpha)\},\]
where $\chi^2_1(\alpha)$ is the $1-\alpha$ quantile of the $\chi_1^2$ distribution. The bandwidth $h$ can be chosen as the optimal bandwidth for estimating $r(x_0)$. In particular, cross-validation can be used to choose a data-dependent bandwidth. In general, as explained above, the constructed confidence interval is valid only for $\mu_0$ instead of $r(x_0)$, while the latter is more commonly the focus of inferences. Part (b) of the theorem shows that the problem can be solved with undersmoothing. But in practice, how to choose the bandwidth is at issue and the asymptotic theory does not provide a solution. Besides, using a smaller bandwidth makes more severe the data sparsity problem which makes estimation of $r(x_0)$ unstable. More discussions and our solution is presented in the next subsection.

The numerical problem is in general easy to tackle since it only involves one-dimensional root-finding. We use Brent's algorithm in our implementation which is faster than the simple bisection algorithm. 

\subsection{Bias-corrected empirical likelihood}\label{sec:bias}
To correct the bias when constructing the confidence interval for $r(x_0)$, we can shift the interval according to the estimated magnitude of the bias. In the proof of asymptotic normality of $\hat{r}(x_0)$ in \cite{masry05}, the bias is $E\hat{r}(x_0)-r(x)=[EKY-r(x_0)EK]/EK\cdot (1+o(1))$, which is in general difficult to estimate. For i.i.d. data, \cite{ferraty07} made the following definition
\[f(s)=E( r(X)-r(x_0)|d(X,x_0)=s),\] 
and imposed the assumption that the derivative $f'(0)$ exists. Under this assumption, the bias is shown to be $f'(0)hM_0/M_1\cdot(1+o(1))$, with $M_0=K(1)-\int_0^1(sK(s))'\tau(s)ds$ and $M_1=K(1)-\int_0^1K'(s)\tau(s)ds$, where the definition of $\tau$ can be found in the Appendix. Because of the difficulty in estimating $\tau$, to compute explicitly $M_0$ and $M_1$ the author used the uniform kernel $K(s)=I_{[0,1]}(s)$. In general, implicit in their calculation is the fact that $M_0=\lim_{h\rightarrow 0} E(d(X,x_0)K)/h\phi(h)$ and $M_1=\lim_{h\rightarrow 0}EK/\phi(h)$ and estimators for $M_0$ and $M_1$ can be easily obtained using sample averages based on this characterization. Nevertheless, in either case, estimating $f'(0)$ looks much more intimidating. 

In this work, we propose to use the bias-corrected estimating equation
\[\sum_i K_i(Y_i-\hat{r}(X_i)+\hat{r}(x_0)-\mu)=0\]
and the correponding empirical likelihood ratio
\[LR^*_n(\mu)=\max\left\{\prod_{i=1}^n (np_i)|\,p_i\ge 0, \sum_{i=1}^np_i=1, \sum_{i=1}^np_iK_i(Y_i-\hat{r}(X_i)+\hat{r}(x_0)-\mu)=0\right\}.\]

The idea is that $K_i(Y_i-\hat{r}(X_i)+\hat{r}(x_0)-\mu)$ will be close to $K_i(Y_i-r(X_i)+r(x_0)-\mu)=K_i(\epsilon_i+r(x_0)-\mu)$, which is obviously an unbiased estimating equation for $r(x_0)$. Formally, we have the following result that shows the bias-corrected empirical likelihood ratio leads to valid confidence interval without the need for undersmoothing.
\begin{theorem}\label{th:2}
Suppose that conditions (c0)-(c9) in the Appendix hold, then 
 $lr_n^*(r(x_0))=-2\log LR_n^*(r(x_0))\rightarrow \chi_1^2$ in distribution.
\end{theorem}

Finally, we mention that the same bias correction can be used in confidence interval constructed with normal approximation theory, with the interval centered at the solution to the bias-corrected estimating equation $\sum_i K_i(Y_i-\hat{r}(X_i)+\hat{r}(x_0)-\mu)=0$ instead of $\hat{r}(x_0)$.

\subsection{Inferences for partially linear models}
Consider the semi-functional partially linear model \citep{engle86,perez06}
\[Y_i=Z_i^T\beta+r(X_i)+\epsilon_i, i=1,\ldots, n,\]
where $Z_i=(Z_{i1},\ldots,Z_{ip})$ is the $p$-dimensional covariate and $\beta=(\beta_1,\ldots,\beta_p)$ is the coefficient for the linear part. We will use $\beta_0$ to denote the true parameter in the following. As before, for the nonparametric part, the covariate $X$ is of functional nature. The estimation of this model is based on the following profiling approach. For a given $\beta$,  the nonparametric part is estimated by 
\[\hat{r}(x,\beta)=\frac{\sum_iK_i(Y_i-Z_i^T\beta)}{\sum_i K_i},\]
where $K_i=K(d(X_i,x)/h)$. Using this definition, the linear coefficient $\beta_0$ is estimated by the profile least square
\[\hat{\beta}=\arg\min_{\beta}\sum_i(Y_i-Z_i^T\beta-\hat{r}(X_i,\beta))^2.\]
After $\hat{\beta}$ is obtained, we estimate the nonparametric part by $\hat{r}(x)=\hat{r}(x,\hat{\beta})$.

The paper \cite{perez06} studied the i.i.d. case and showed that $\sqrt{n}(\hat{\beta}-\beta)$ has an asymptotically normal distribution under mild assumptions. Instead of repeating their assumptions and extending their arguments for strongly mixing sequences, we directly impose the assumption that $||\hat{\beta}-\beta||=O_p(n^{-1/2})$ for simplicity. 

Using a plug-in approach, the empirical likelihood ratio can be defined as
\[\widetilde{LR}_n(\mu)=\max\left\{\prod_{i=1}^n (np_i)|\,p_i\ge 0, \sum_{i=1}^np_i=1, \sum_{i=1}^np_iK_i(Y_i-Z_i^T\hat{\beta}-\mu)=0\right\},\]
and the bias-corrected version is
\[\widetilde{LR}^*_n(\mu)=\max\left\{\prod_{i=1}^n (np_i)|\,p_i\ge 0, \sum_{i=1}^np_i=1, \sum_{i=1}^np_iK_i(Y_i-Z_i^T\hat{\beta}-\hat{r}(X_i)+\hat{r}(x_0)-\mu)=0\right\}.\]
\begin{theorem}\label{th:3}
Suppose that conditions (c0)-(c9) in the Appendix hold, and $||\hat{\beta}-\beta_0||=O_p(n^{-1/2})$, then 
 $\widetilde{lr}^*_n(r(x_0))=-2\log \widetilde{LR}_n^*(r(x_0))\rightarrow \chi_1^2$ in distribution.
\end{theorem}
Most previous studies on empirical likelihood inference for partially linear models \citep{shi00,lu09,chen08} focused on constructing confidence regions for the linear coefficient $\beta$. For our functional model, similar confidence regions can be developed for linear coefficients without much difficulty, but we choose to focus on the nonparametric part to better match the main theme of the paper.

\section{Numerical results}
\subsection{Simulation data}
We first use a simulated example to show the performance of the empirical likelihood based confidence interval and compare it to that based on normal approximation. We simulate i.i.d. samples from the nonparametric functional model using the following mean function
\[r(x)=\int_{-1}^1|x'(t)|(1-\cos(\pi t))dt.\]
The random covariate curves are simulated from
\[X(t)=\sin(\omega t)+(a+2\pi)t+b,\; \omega\sim Unif(0,2\pi),\; a,b\sim Unif(0,1),\]
and the noise $\epsilon_i$ is simulated from a $N(0,\sigma^2)$ distribution. This example is the same as that used in \cite{ferraty07} to illustrate bootstrap bandwidth selection.

We use $n=200$ and $n=500$ as well as $\sigma^2=0.5$ and $\sigma^2=2$ resulting in a combination of four scenarios. We use the quadratic kernel $K(s)=(1-s^2), 0\le s\le 1$ and the $L^2$ distance between the first derivatives of the curves as the semi-metric since the true regression function directly depends on the first derivative. The bandwidth is chosen using cross-validation, taking advantage of the \textbf{npfda} R software publicly available online (\url{http://www.lsp.ups-tlse.fr/staph/npfda}). 

For each simulation scenario, we randomly generated $100$ testing curves and we constructed $95\%$ confidence intervals for them using both empirical likelihood and normal approximation, with bias either ignored or corrected based on the approach presented in section \ref{sec:bias}. The whole simulation process is repeated $50$ times. The percentage of times it covers the true $r(x)$ as well as the average interval length is calculated. As shown by \cite{ferraty07}, the asymptotic variance of $\hat{r}(x_0)$ is $\sigma^2M_2/(n\phi(h)M_1^2)$, with relevant constants $M_1$ and $M_2$ defined in the Appendix. For the normal approximation approach, we need to estimate the variance as well as the constants $M_1$ and $M_2$. As shown in the Appendix we have $M_1=\lim_{h\rightarrow 0}EK/\phi(h)$ and $M_2=\lim_{h\rightarrow 0} EK^2/\phi(h)$. Thus $M_2/(n\phi(h)M_1^2)$ can be estimated by the sample version $\sum_iK_i^2/(\sum_iK_i)^2$. The noise variance $\sigma^2$ will be estimated by the mean residual $\hat{\sigma}^2=(Y_i-\hat{r}(X_i))^2/n$.

The simulation results shown in Table \ref{tab:sim} demonstrated the superiority of empirical likelihood based intervals. For all cases, the empirical likelihood method produces better coverage and shorter intervals compared to the normal approximation based method. We also observe that the bias-corrected intervals give improved coverage accuracy. 
\begin{table}
\centering
\begin{tabular}{cccccc}
 n & $\sigma^2$ & EL &Normal& Corrected EL &Corrected Normal\\
\hline
200&  0.5        &0.912& 0.887& 0.933&0.901\\
   &             &(0.85)&(0.90)&(0.87)&\\
   &  2.0        &0.914&0.879&0.930&0.895\\
   &             &(1.17)&(1.28)&(1.15)&\\
500&  0.5        &0.920&0.891&0.943&0.911\\
   &             &(0.62)&(0.72)&(0.66)&\\
   &  2.0        &0.921&0.890&0.946&0.905\\ 
   &             &(0.91)&(1.00)&(0.93)&\\
\end{tabular}
\caption{\label{tab:sim} Simulation results for the constructed $95\%$ confidence interval. The numbers shown are the coverage accuracy and the average interval lengths (numbers in the brackets). }
\end{table}
\subsection{Real data}
In this subsection we use two real datasets to illustrate the construction of confidence intervals for functional data in nonparametric regression. First, our spectrometrics dataset contains as covariates 215 spectra of light absorbance as functions of the wavelength, for 215 pieces of finely chopped meat. The dependent variable is the percentage of fat in each piece of meat. Besides, we use the protein and moisture content as two covariates of the linear part in a partially linear model since a previous study \citep{perez06} shows these additional covariates give better prediction performance. The first 165 samples are used as training data, the rest 50 samples are used as testing data and $95\%$ confidence intervals are constructed for the testing data. The semi-metric used is the $L^2$ metric on the first derivatives of the spectra curves, which gives best prediction performance on this dataset as shown in \cite{perez06}. Since the convergence rate of the linear part is faster than the nonparametric part, it seems reasonable as a rough approximation to shift the confidence interval for $r(x)$ by $z^T\hat{\beta}$ and treat it as a confidence interval for the entire regression function $z^T\beta+r(x)$. 

Our El Ni\~{n}o time series dataset records the monthly sea surface temperature from June, 1950 up to May, 2004 (648 months). 
We use the first 53 years as training and the final year temperature as our testing data. The $j$-th month temperature in a certain year is predicted from a nonparametric regression model using the whole previous year observations treated as a curve. Thus the prediction for each month in the future is based on a different regression model. For this dataset, we use the PCA semi-metric \citep{ferraty06} with the first four principal components.

\begin{figure}
\centering
\includegraphics[width=8cm]{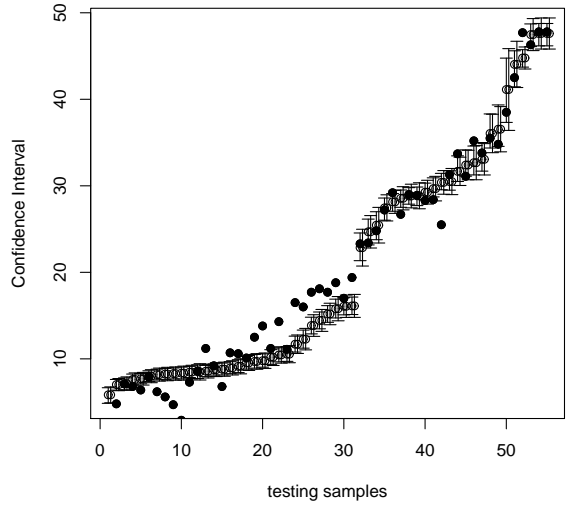}\\
(a)\\
\includegraphics[width=8cm]{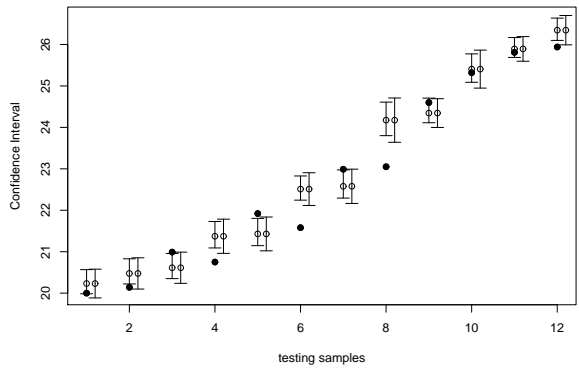}\\
(b)
\caption{\label{fig:real} Two-sided 95\% Confidence intervals constructed using both empirical likelihood and normal approximation on the testing samples of (a) the spectrometrics dataset; (b) the El Ni\~{n}o dataset. The intervals constructed using normal approximation are shifted slightly to the right for visualization. The solid circles denote the observed responses.}
\end{figure}

For both datasets, the constructed confidence intervals for the testing data are shown in Figure \ref{fig:real} using empirical likelihood as well as normal approximation, with the bias-corrected version. For better visualization, the testing samples are sorted according to the estimated responses. Consistent with our simulations, the normal intervals are generally longer, with an average length of 2.69 and 0.78 for the two datasets respectively, compared to the average length of 2.17 and 0.62 for the empirical likelihood intervals. In these two datasets, although we have no way of assessing the coverage rate of the constructed intervals, we believe the empirical likelihood intervals are better based on our previous simulation results. The mean squared error (MSE) for the bias-corrected estimator on the testing data are $3.78$ and $4.04$ in the two datasets, which compares favorably with MSE for the uncorrected estimator, $5.36$ and $4.49$ respectively.

\section{Conclusions}
The construction of confidence intervals should accompany any statistical analysis where a point estimate is obtained. In this paper, we propose to use empirical likelihood for such purposes for the nonparametric functional data model. The popularity of empirical likelihood derives from its data-dependent shape of the constructed confidence regions, resulting in typically better finite sample performance. The asymptotic property of the empirical likelihood ratio is demonstrated and we also propose a bias correction method that avoids undersmoothing or direct estimation of the bias. We also extend our result to semi-functional partially linear models where the confidence interval for the nonparametric part can be constructed. Our asymptotic results are presented in the general context of strongly mixing data sequence and thus applicable to time series datasets.

Similar results can be obtained when the empirical likelihood is replaced by other distance functions. A particularly simple alternative is the Euclidean log likelihood ratio $\sum_i(np_i-1)^2$ \citep{owen01}. In this case, one can show the Lagrange multiplier $\lambda$ can be eliminated to obtain $\sum_i(np_i-1)^2=(\sum_iK_i(Y_i-\mu))^2/\sum_i[(K_i(Y_i-\mu)-\bar{KY}+\bar{K}\mu)^2]$, where $\bar{KY}=\sum_iK_iY_i/n$ and $\bar{K}=\sum_iK_i/n$. Thus the Euclidean likelihood has some advantage over the empirical likelihood in terms of computations.

\section*{Appendix: Proofs}
We first list some conditions assumed in the theorems:
\begin{itemize}
\item[(c0)]$P(LR_n(\mu)=0)\rightarrow 0$, and similarly for $LR_n^*$, $\widetilde{LR}_n$ and $\widetilde{LR}_n^*$, for either $\mu=\mu_0$ or $\mu=r(x_0)$.
\item[(c1)]The kernel $K$ is supported on $[0, 1]$, and its derivative $K'$ exists on $[0, 1]$. Either $K $ is bounded and bounded away from zero on $[0, 1]$,
    or, $K'<0$ and bounded way from zero on $[0,1]$, and for $\delta>0$ small enough, $\int_0^\delta\phi(h)dh>C\delta\phi(\delta)$ for some constant $C>0$.

\item[(c2)] The regression function $r(x)$ satisfies the Lipschitz condition: $|r(x_1)-r(x_2)|\le Cd(x_1,x_2)^\alpha$. The variance function $\sigma^2(x)$ is continuous in a neighborhood of $x_0$. 
\item[(c3)] The distribution of the random covariate $X$ and $Z$ are both compactly supported, and the third conditional moment for $Y$ satisfies $E(|Y|^3|X=x)<C<\infty$ in a neighborhood of $x_0$.

\item[(c4)] For all $0\le s\le 1$, $\lim_{h\rightarrow 0}\phi(hs)/\phi(h)\rightarrow\tau(s)$ for some function $\tau$.
\item[(c5)] $\psi(h)/\phi(h)^2$ is bounded.
\item[(c6)] The sequence $(Y_i,X_i)$ is $\alpha$-mixing with mixing coefficients $\alpha(\cdot)$ satisfying $\sum_{l=1}^\infty l^\gamma[\alpha(l)]^{1-2/\nu}<\infty$ for some $\nu>2$ and $\gamma>1-2/\nu$.
\item[(c7)] There exists a sequence $v_n$ satisfying $v_n\rightarrow\infty$, $v_n=o((n\phi(h))^{1/2})$ and $(n/\phi(h))^{1/2}\alpha(v_n)\rightarrow 0$.
\item[(c8)] $h\rightarrow 0, n\phi(h)\rightarrow \infty, nh^{2\alpha}\phi(h)=O(1)$.
\item[(c9)] The kernel $K$ is continuous on $(0,\infty)$.
\end{itemize}
Condition (c0) is the basic assumption for empirical likelihood. For example, in nonparametric regression, it is equivalent to saying that there exists two samples $i$ and $j$ such that, $K_i>0, K_j>0, Y_i>\mu$ and $Y_j<\mu$. This of course can be guaranteed with mild assumptions on the distribution of error. We directly put it down as one assumption to avoid verification for different cases. Condition (c1) guarantees that $EK(d(X,x_0)/h)$ has the same asymptotic order as $\phi(h)$ (see Lemma 4.3 and 4.4 in \cite{ferraty06}). The Lipschitz assumption directly determines the order of the bias of the Nadaraya-Watson estimator. The assumption on the support of $X$ and $Z$ is for technical reasons to simplify several arguments in the proofs, which can be relaxed with more careful analysis. Condition (c4) is also stated in \cite{ferraty07} and is used to get explicit expression for asymptotic bias and variance. Conditions (c5) and (c6) are the standard conditions used to show that the sequence is only weakly dependent so that the asymptotic properties of various statistics is similar to the i.i.d. case. The existence of sequence $v_n$ in condition (c7) is used for the standard big-block small-block argument for weakly dependent data. As is well-known for the Nadaraya-Watson estimator the bias is of order $O(h^\alpha)$ and the variance is of order $O((n\phi(h))^{-1})$. Thus the asymptotically optimal bandwidth satisfies condition (c8) which provides the correct trade-off between bias and variance. Although this is not required for all of our results, it is assumed throughout for simplicity. Finally, condition (c9) will only be used in the proof of of the Wilks's theorem for the bias-corrected empirical likelihood, where the continuity of kernel is required to show the bias can be correctly eliminated. In particular, since we assume $K$ is supported on $[0,1]$, this implies $K(1)=0$. \\

We first introduce the following constants:
\[M_1=K(1)-\int_0^1(sK(s))'\tau(s)ds,\]
\[M_2=K^2(1)-\int_0^1(K^2)'(s)\tau(s)ds.\]
These constants are defined in \cite{ferraty07}, and actually from their calculations, we find $M_1=\lim_{h\rightarrow 0}EK/\phi(h)$ and $M_2=\lim_{h\rightarrow 0}EK^2/\phi(h)$.\\

\noindent\textit{Proof of Theorem \ref{th:1}.}
(a) The general approach to show the convergence in distribution of the empirical likelihood ratio is laid out in \cite{hjort09}. In particular, we only need to verify their assumptions (A1)-(A3), while assumption (A0) in their paper is directly assumed in (c0). Thus our proof will be split into three steps.

Step 1. We show 
\[\frac{\sum_iK_i(Y_i-\mu_0)}{\sqrt{nEK}}\rightarrow N(0,\frac{M_2}{M_1}\sigma^2(x_0)) \mbox{ in distribution}.\]

Since we deal with dependent data here, the proof of central limit theorem is more involved. The main ideas used include big-block small-block trick, bounding covariances using mixing coefficients, and finally showing asymptotic normality from convergence of characteristic functions. We omit the details here since they are almost identical to the proof in \cite{masry05} in showing the asymptotic normality of $\hat{r}(x_0)$. We only calculate the asymptotic mean and variance of $K(Y-\mu_0)/\sqrt{EK}$ which gives us the expression in the asymptotic normal distribution above.

The mean is obviously zero and the variance is calculated as 
\begin{eqnarray*}
Var(K(Y-\mu_0)/\sqrt{EK})&=&\frac{1}{EK}EK^2(Y-\mu_0)^2=\frac{1}{EK}(EK^2Y^2-2\mu_0EK^2Y+\mu_0^2EK^2).
\end{eqnarray*}
We have 
\[EK^2Y^2=E(r^2(X)K^2)+E(\sigma^2(X)K^2)=(\sigma^2(x_0)+r^2(x_0)+o(1))EK^2,\] 
by the continuity of $r(x)$ and $\sigma^2(x)$. Similarly, $EK^2Y=(r(x_0)+o(1))EK^2$. Thus the variance is
\begin{eqnarray*}
&&Var(K(Y-\mu_0)/\sqrt{EK})\\
&=&\{(\sigma^2(x_0)+r^2(x_0)+o(1))EK^2-2\mu_0r(x_0)EK^2+\mu_0^2EK^2\}/EK\\
&=&(\sigma^2(x_0)+(r(x_0)-\mu_0)^2+o(1))EK^2/EK\\
&\rightarrow &\sigma^2(x_0)M_2/M_1.
\end{eqnarray*}

Step 2. We show $\sum_i\frac{K_i^2(Y_i-\mu_0)^2}{nEK}\rightarrow \sigma^2(x_0)M_2/M_1$ in probability.
As calculated in Step 1, $E[K^2(Y-\mu_0)^2/EK]\rightarrow \sigma^2(x_0)M_2/M_1$ and the result follows from the ergodic theorem.

Step 3. We show $\max_{1\le i\le n} \left|\frac{K_i(Y_i-\mu_0)}{\sqrt{nEK}}\right|\rightarrow 0$ in probability.

We have by the union bound and Markov inequality that 
\begin{eqnarray*}
&&P(\max_i|K_i(Y_i-\mu_0)|>\delta\sqrt{nEK})\\
&\le& nP(|K(Y-\mu_0)|^3>\delta^3(nEK)^{3/2})\\
&=&O(\frac{1}{\sqrt{nEK}}),
\end{eqnarray*}
using condition (c3).

Now the conditions (A1)-(A3) in \cite{hjort09} are verified with the asymptotic variance in Step 1 same as the converged constant in Step 2, and part (a) in the theorem is proved.

(b) The proof is very similar to (a). In particular, the calculations in Step 2 and 3 above remain exactly the same with $\mu_0$ replaced by $r(x_0)$. For Step 1, we have
\[\frac{\sum_iK_i(Y_i-r(x_0))}{\sqrt{nEK}}=\frac{\sum_iK_i(Y_i-\mu_0)}{\sqrt{nEK}}+\frac{\sum_iK_i(\mu_0-r(x_0))}{\sqrt{nEK}}.\]

We note that the term $\mu_0-r(x_0)$ is same as the bias term in estimating $r(x_0)$ and thus $|\mu_0-r(x_0)|=O(h^\alpha)=o((n\phi(h))^{-1/2})$ by the results in \cite{ferraty04} and the assumption $nh^{2\alpha}\phi(h)\rightarrow 0$. The following Lemma \ref{lem:1} immediately implies that $\frac{\sum_iK_i(Y_i-r(x))}{\sqrt{nEK}}$ and $\frac{\sum_iK_i(Y_i-\mu_0)}{\sqrt{nEK}}$ have the same asymptotic distribution.\\

\begin{lemma}\label{lem:1} 
For any random sequence $a_i, 1\le i\le n$, we have $\frac{\sum_iK_ia_i}{\sqrt{nEK}}=o_p(1)$ if $\max_i|a_i|=o_p((n\phi(h))^{-1/2})$.
\end{lemma}
\textit{Proof of Lemma \ref{lem:1}.} 
$\frac{\sum_iK_ia_i}{nEK}\le \frac{\sum_iK_i\max_i|a_i|}{nEK}=O_p(\max_i|a_i|)=o_p((n\phi(h))^{-1/2})$ since $\sum_i K_i/(nEK)\rightarrow 1$ by the ergodic theorem. Combine this with the fact that $EK$ is of the same order as $\phi(h)$ to get the result.\\

\noindent\textit{Proof of Theorem \ref{th:2}.}
As in Theorem 1, we split the proof into three steps. Step 2 and 3 still remain the same with $K_i(Y_i-\mu_0)$ replaced by $K_i(Y_i-r(x_0)-\hat{r}(X_i)+\hat{r}(x_0))$ using the fact that $|\hat{r}(X_i)-\hat{r}(x_0)|=o_p(1)$ when $K_i>0$. We only need to replace Step 1 to show 
that $\sum_iK_i(Y_i-r(x_0)-\hat{r}(X_i)+\hat{r}(x_0))/\sqrt{nEK}\rightarrow N(0,\sigma^2(x_0)M_2/M_1)$.

To simplify notation, we set $K_{ij}=K(d(X_i,X_j)/h)$ and $w_{ij}=K_{ij}/\sum_jK_{ij}$. We note that even though $w_{ij}\neq w_{ji}$, we have $w_{ij}=w_{ji}(1+o_p(1))$. Similarly, let $w_i=K_i/\sum_jK_j=K(d(X_i,x_0)/h)/\sum_jK(d(X_j,x_0)/h)$.
We have the decomposition 
\begin{eqnarray*}
&&\sum_iK_i(Y_i-r(x_0)-\hat{r}(X_i)+\hat{r}(x_0))\\
&=&\sum_iK_i(Y_i-r(X_i))+\sum_iK_i(r(X_i)-r(x_0)-\hat{r}(X_i)+\hat{r}(x_0))\\
&=&\sum_iK_i(Y_i-r(X_i))-\sum_iK_i(\sum_jw_{ij}Y_j-\sum_jw_jY_j-r(X_i)+r(x_0))\\
&=&\sum_iK_i(Y_i-r(X_i))-\sum_iK_i(\sum_jw_{ij}\epsilon_j-\sum_jw_j\epsilon_j)\\
&&\;\;\;-\sum_iK_i(\sum_jw_{ij}r(X_j)-\sum_jw_jr(X_j)-r(X_i)+r(x_0))\\
&=&(I)-(II)-(III).
\end{eqnarray*}
Dealing with $(I)$, similar to the proof of Step 1 in Theorem \ref{th:1}, we have $\sum_iK_i(Y_i-r(X_i))/\sqrt{nEK}$ converges to $N(0,\sigma^2(x_0)M_2/M_1)$ in distribution. And we only need to show $(II)=o_p(\sqrt{nEK})$ and $(III)=o_p(\sqrt{nEK})$. This is done in the following two lemmas for better readability.

\begin{lemma} \label{lem:2}
$\sum_iK_i(\sum_jw_{ij}\epsilon_j-\sum_jw_j\epsilon_j)=o_p(\sqrt{nEK})$.
\end{lemma}
\textit{Proof of Lemma \ref{lem:2}.} Denote the expression on the left hand side by $g_n$. We bound the variance of $g_n$ conditional on $\{X_i,i=1,\ldots,n\}$ as
\[Var(g_n|\{X_i\})=\sum_j[\sum_i(K_iw_{ij})-K_j]^2.\]
To see more clearly the order of each term in the sum, for a fixed $j$, we set $m(x):=K(d(x,x_0)/h)$. Then $\sum_i(K_iw_{ij})-K_j=\sum_i[(w_{ji}(1+o_p(1)))m(X_i)]-m(X_j)=o_p(1)$ due to the continuity of $m$.

Thus we have $E(g_n^2|\{X_i\})=Var(g_n|\{X_i\})=o_p(nEK)$ which implies $g_n=o_p(\sqrt{nEK})$.

\begin{lemma} \label{lem:3}
$\sum_iK_i(\sum_jw_{ij}r(X_j)-\sum_jw_jr(X_j)-r(X_i)+r(x_0))=o_p(\sqrt{nEK})$.
\end{lemma}
\textit{Proof of Lemma \ref{lem:3}.} The left hand side is equal to 
\[\sum_j\{\sum_iK_i(1+o(1))w_{ji}(r(X_j)-r(X_i))-K_j(r(X_j)-r(x_0))\}\]
Fixing any $j$ and setting $m(x):=K(d(X_j,x)/h)(r(X_j)-r(x))/h^\alpha$. Note $m$ is bounded due to the Lipschitz condition (c2) for $r$. Each term above inside the sum over $j$ is rewritten as $\{\sum_iK_i(1+o_p(1))m(X_i)-m(x_0)\}h^\alpha=o_p(h^\alpha)=o_p(1/\sqrt{nEK})$ by the same argument as in Lemma \ref{lem:2} as well as condition (c8). Thus the left hand side of the expression in the statement of the lemma is $o(\sqrt{nEK}).$\\

\noindent\textit{Proof of Theorem \ref{th:3}}.
With the partially linear model
\[Y_i-Z_i^T\beta_0=r(X_i)+\epsilon_i,\]
we see that the empirical likelihood ratio in Theorem \ref{th:3} with true $\beta_0$ replacing $\hat{\beta}$ will have the desired convergence property by apply Theorem \ref{th:2} to $Y_i-Z_i^T\beta_0$ instead of $Y_i$. Using the boundedness of $Z_i$ and the assumption that $||\hat{\beta}-\beta_0||=O_p(n^{-1/2})$, together with Lemma \ref{lem:1}, we can still verify the three steps in the proofs of Theorems \ref{th:1} and \ref{th:2}, and thus the empirical likelihood ratio defined using either $\beta_0$ or $\hat{\beta}$ are asymptotically equivalent.

\bibliographystyle{elsarticle-harv}
\bibliography{papers.txt,books.txt}

\end{document}